\documentstyle[prl,aps,amssymb,mathrsfs,epsfig]{revtex}
\begin{document}
\draft

\def\overlay#1#2{\setbox0=\hbox{#1}\setbox1=\hbox to \wd0{\hss #2\hss}#1%
\hskip
-2\wd0\copy1}
\twocolumn[
\hsize\textwidth\columnwidth\hsize\csname@twocolumnfalse\endcsname

%\twocolumn[
\title{Quantized R\"ontgen Effect in Bose--Einstein Condensates}
\author{U.\ Leonhardt$^1$ and P.\ Piwnicki$^{1,2}$}
\address{~$^1$Physics Department, Royal Institute of Technology (KTH),
Lindstedtsv\"agen 24, S-10044 Stockholm, Sweden}
\address{~$^2$Abteilung f\"ur Quantenphysik, Universit\"at Ulm,
D--89069 Ulm, Germany}
\maketitle
%\mediumtext
\begin{abstract}
A classical dielectric moving in a charged capacitor can create
a magnetic field (R\"ontgen effect).
A quantum dielectric, however, will not produce a magnetization,
except at vortices.
The magnetic field outside the quantum dielectric appears as the 
field of quantized monopoles.
\end{abstract}
\date{today}
%]
\vskip2pc]
\narrowtext
%\narrowtext
%%%%
Bose--Einstein condensation is believed to be at the heart of 
superfluidity and superconductivity \cite{Tilley}. 
Formulated in the most elementary model,
a large number of either neutral atoms or electrically charged Cooper pairs
condense and constitute a macroscopic wave function. 
Flowing Cooper pairs form electric currents that in turn 
act on magnetic fields, and the magnetic properties of superconductors
cause the phenomenon of electric superconductivity itself \cite{LL8}. 
On the other hand, superfluids, i.e. Bose--condensed atoms, 
are electrically neutral, yet they are polarizable 
and form a dielectric medium.

Can quantum dielectrics generate magnetic fields? 
Imagine the setup depicted in Fig.\ 1.
\begin{figure}[htbp]
  \begin{center}
    \epsfig{file=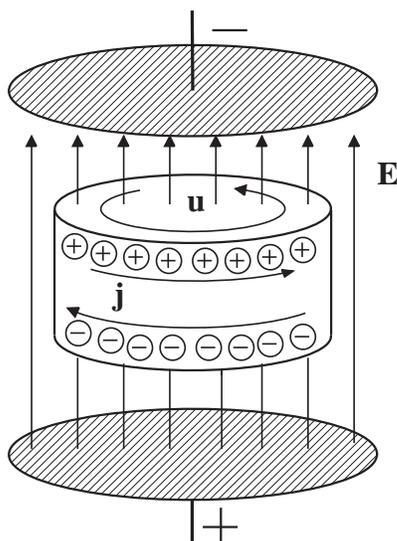,width=.3\textwidth}
    \caption{Sketch of R\"ontgen's experiment.
     A dielectric moves with the velocity profile ${\bf u}$
     in a charged capacitor.
     The co--moving surface charges of the polarized medium 
     form a current ${\bf j}$
     and thus generate a magnetic field.
     }
    \label{fig:figure1}
  \end{center}
\end{figure}
\noindent
A capacitor polarizes the dielectric layer between the plates. 
When the medium is moving, co--moving surface charges appear as 
electric currents and produce a magnetic field. 
In 1888 W.C.\ R\"ontgen \cite{Roentgen} 
observed the effect for the first time (before he discovered X-rays).
R\"ontgen employed a rotating glass disc (and also a rubber disc) as a moving
dielectric medium in a charged capacitor, and he noticed a feeble deflection 
of a magnetic needle. 
More recently the interest in R\"ontgen's effect \cite{Optics} has
revived \cite{Wilkens,China,Discussion,Experiment}, 
because the effect gives rise to a novel topological phase of 
neutral atoms.

What happens when R\"ontgen's glass disc is replaced by a quantum dielectric
(a Bose--Einstein condensate)?
We show in this paper that only vortices of the quantum liquid 
can generate a magnetization. 
Moreover, the magnetic field outside a thin quantum--dielectric layer 
appears as the field of a set of magnetic monopoles 
with magnetic charges that sit in the vortex cores. 
The monopoles turn out to be quantized \cite{Dirac} with a magnetic flux of
%%%%%%%%%%%%%%%%%%%%%%%%%%%%%%%%%%%%%%%%%%%%%%%%%%%%%%%%%%%%%
\begin{equation}\label{result}
\Phi_n=\frac{hn}{mc^2}\,\chi\, U \,.
\end{equation}
%%%%%%%%%%%%%%%%%%%%%%%%%%%%%%%%%%%%%%%%%%%%%%%%%%%%%%%%%%%%%
Here $U$ denotes the applied voltage, 
$\chi$ is the susceptibility,
$m$ is the atomic mass, 
$c$ denotes as usual the speed of light, and 
$hn$ is an integer multiple of Planck's constant $h=2 \pi \hbar$. 
The quantized R\"ontgen effect of a single vortex is very small, 
yet modern ``magnetic needles'' (SQUIDs) might be able to detect it. 
Let us present our case beginning with a general theory of 
condensed atoms that move in electromagnetic fields. 
Then we turn to the quantum version of R\"ontgen's experiment.

{\it Moving atoms in electromagnetic fields.}---
Any magnetic effect has its root in relativity. 
Let us first consider a classical atom that moves with velocity
$\mathbf v$ in a given electromagnetic field. 
The neutral yet polarizable atom responds to the electric field 
in its restframe. 
According to relativity, this field contains an electric and a 
magnetic component of the electromagnetic field in the laboratory frame. 
In lowest order of $v/c$ we use the mechanical Lagrangian \cite{China}
(in SI units)
%%%%%%%%%%%%%%%%%%%%%%%%%%%%%%%%%%%%%%%%%%%%%%%%%%%%%%%%%%%%%
\begin{equation}
\label{la}
L_A=\frac{m}{2}v^2+\frac{\alpha}{2}E^2-
\alpha {\bf v}({\bf E}\times{\bf B})\,.
\end{equation}
%%%%%%%%%%%%%%%%%%%%%%%%%%%%%%%%%%%%%%%%%%%%%%%%%%%%%%%%%%%%%
The constant $\alpha$ denotes the electrical polarizability of the atom. 
Due to relativity 
the Poynting vector ${\bf E}\times{\bf B}$ of the electromagnetic
field couples to the atom's motion in precisely the same way as the
vector potential couples to a charged particle. 

To study the motion of an atomic de--Broglie wave $\psi$ in 
an electromagnetic field, we follow the canonical procedure. 
First, we find the momentum ${\bf p}$ and the Hamiltonian $H$ 
of the classical atom,
%%%%%%%%%%%%%%%%%%%%%%%%%%%%%%%%%%%%%%%%%%%%%%%%%%%%%%%%%%%%%
\begin{eqnarray}
{\bf p}&=&\frac{\partial L}{\partial {\bf v}}
=m {\bf v} - \alpha ({\bf E}\times{\bf B}) \,,
\nonumber\\
H&=&{\bf p v}-L=\frac{1}{2m}({\bf p}+ \alpha {\bf E}\times{\bf B})^2-
\frac{\alpha}{2}E^2\,,
\end{eqnarray}
%%%%%%%%%%%%%%%%%%%%%%%%%%%%%%%%%%%%%%%%%%%%%%%%%%%%%%%%%%%%%
and then we write down the Schr\"odinger equation
%%%%%%%%%%%%%%%%%%%%%%%%%%%%%%%%%%%%%%%%%%%%%%%%%%%%%%%%%%%%%
\begin{equation}\label{sch}
i\hbar\frac{\partial \psi}{\partial t}
=\frac{1}{2m}(-i\hbar\nabla + \alpha {\bf E}\times{\bf B})^2\psi -
\frac{\alpha}{2}E^2 \psi\,.
\end{equation}
%%%%%%%%%%%%%%%%%%%%%%%%%%%%%%%%%%%%%%%%%%%%%%%%%%%%%%%%%%%%%
In the case of a Bose--Einstein condensate, $|\psi|^2$ describes 
the density of the condensed atoms 
(in mean--field theory \cite{DGPS}). 
The condensate moves with the velocity profile 
%%%%%%%%%%%%%%%%%%%%%%%%%%%%%%%%%%%%%%%%%%%%%%%%%%%%%%%%%%%%%
\begin{equation}\label{u}
{\bf u}=\frac{1}{m}(\hbar \nabla S+ \alpha {\bf E}\times{\bf B})\,, \qquad
\psi=|\psi| e^{iS}\,,
\end{equation}
%%%%%%%%%%%%%%%%%%%%%%%%%%%%%%%%%%%%%%%%%%%%%%%%%%%%%%%%%%%%%
because ${\bf u}$ satisfies the continuity relation
%%%%%%%%%%%%%%%%%%%%%%%%%%%%%%%%%%%%%%%%%%%%%%%%%%%%%%%%%%%%%
\begin{equation}\label{c}
\frac{\partial |\psi|^2}{\partial t}+\nabla(|\psi|^2{\bf u})=0 \,.
\end{equation}
%%%%%%%%%%%%%%%%%%%%%%%%%%%%%%%%%%%%%%%%%%%%%%%%%%%%%%%%%%%%%
Condensed atoms interact by collisions.
In order to form a stable condensate for large numbers of particles,
the atoms must repell each other \cite{DGPS}.
We model atomic collisions  by adding a Gross--Pitaevskii term 
$g |\psi|^2\psi$ \cite{DGPS} with positive $g$ 
to the  right--hand side of the Schr\"odinger equation (\ref{sch}).
The atomic repulsion tends to smooth out density variations
over the healing length $\hbar/\sqrt{2gm |\psi|^2}$ \cite{DGPS}.
To prevent the condensate from spreading out to infinity,
an external potential $V$ must balance out the inter--atomic repulsion.
The potential $V$ models a trap or simply the interactions with the
walls of a container.
Finally, we condense our description of moving atoms in
electromagnetic fields into the Lagrangian density
%%%%%%%%%%%%%%%%%%%%%%%%%%%%%%%%%%%%%%%%%%%%%%%%%%%%%%%%%%%%%
\begin{eqnarray}
{\mathscr L}_A&=&
\frac{i\hbar}{2}
(\psi^{\ast}\dot{\psi}-\dot{\psi^{\ast}}\psi)+
\left(\frac{\alpha}{2}E^2 - \frac{g}{2}\,|\psi|^2 - V
\right)|\psi|^2
\nonumber\\
&-&
\frac{1}{2m}(i\hbar\nabla + \alpha {\bf E}\times{\bf B})\psi^{\ast}
\cdot
(-i\hbar\nabla + \alpha {\bf E}\times{\bf B})\psi
\,.
\end{eqnarray}
%%%%%%%%%%%%%%%%%%%%%%%%%%%%%%%%%%%%%%%%%%%%%%%%%%%%%%%%%%%%%
One verifies easily that the Euler--Lagrange equations of 
${\mathscr L}_A$ lead to the Gross--Pitaevskii equation 
(including electromagnetic and external interactions).

So far we have described atoms that move in given electromagnetic fields.
For understanding the condensate as a dielectric medium we must study the
effect of the atoms on the fields. 
For this we add to ${\mathscr L}_A$ the Lagrangian density 
${\mathscr L}_F$ of the free electromagnetic field (in SI units),
and we arrive at the total Lagrangian
%%%%%%%%%%%%%%%%%%%%%%%%%%%%%%%%%%%%%%%%%%%%%%%%%%%%%%%%%%%%%
\begin{eqnarray}
{\mathscr L}&=&{\mathscr L}_F+{\mathscr L}_A\,,
\\
{\mathscr L}_F&=&\frac{\varepsilon_0}{2}(E^2-c^2 B^2)\,,
\nonumber\\ 
{\bf E}&=&-(\dot{\bf A} + \nabla U), \quad {\bf B}=\nabla\times {\bf A}\,,
\end{eqnarray}
%%%%%%%%%%%%%%%%%%%%%%%%%%%%%%%%%%%%%%%%%%%%%%%%%%%%%%%%%%%%%
where ${\bf A}$ denotes the vector potential.
Then we minimize the total action $\int{\mathscr L} \,d^4 x$
with respect to $\bf A$ and $U$, and obtain 
%%%%%%%%%%%%%%%%%%%%%%%%%%%%%%%%%%%%%%%%%%%%%%%%%%%%%%%%%%%%%
\begin{eqnarray}
\nabla\times{\bf E}&=&-\frac{\partial {\bf B}}{\partial t}\,, 
\qquad \nabla {\bf B}=0\,, 
\nonumber \\
\nabla\times{\bf H}&=&+\frac{\partial {\bf D}}{\partial t}\,, 
\qquad \nabla {\bf D}=0\,,
\label{maxwell}
\end{eqnarray}
%%%%%%%%%%%%%%%%%%%%%%%%%%%%%%%%%%%%%%%%%%%%%%%%%%%%%%%%%%%%%
with 
%%%%%%%%%%%%%%%%%%%%%%%%%%%%%%%%%%%%%%%%%%%%%%%%%%%%%%%%%%%%%
\begin{eqnarray}
{\bf D}&=&\varepsilon_0(1+\chi){\bf E} +
\varepsilon_0\chi {\bf u}\times {\bf B} \,,
\nonumber\\ 
{\bf H}&=&\varepsilon_0 c^2{\bf B} +
\varepsilon_0\chi {\bf u}\times {\bf  E}
\,,
\label{constitution}
\end{eqnarray}
%%%%%%%%%%%%%%%%%%%%%%%%%%%%%%%%%%%%%%%%%%%%%%%%%%%%%%%%%%%%%
and
%%%%%%%%%%%%%%%%%%%%%%%%%%%%%%%%%%%%%%%%%%%%%%%%%%%%%%%%%%%%%
\begin{equation}
\varepsilon_0\chi=\alpha\,|\psi|^2\,.
\label{suscept}
\end{equation}
%%%%%%%%%%%%%%%%%%%%%%%%%%%%%%%%%%%%%%%%%%%%%%%%%%%%%%%%%%%%%
These are Maxwell's equations for the electromagnetic field
in the presence of a moving dielectric medium
\cite{LandauMotion,Bladel},
written in lowest order of $u/c$.
The medium has a susceptibility $\chi$ given by Eq.\ (\ref{suscept})
and moves with the velocity profile ${\bf u}$ of Eq.\ (\ref{u}).
Consequently, moving atoms form indeed a classical
dielectric medium (like R\"ontgen's glass disc). 
On the other hand, where are the quantum effects?

{\it Quantized R\"ontgen effect.}---
Consider a charged capacitor that contains a movable quantum dielectric,
see Fig.\ 1. 
We will use cylindrical coordinates $(\varrho,\varphi,z)$
to describe the physical situation. 
We require that no motion occurs in $z$ direction and that the
susceptibility $\chi$ (the density $|\psi|^2$) be uniform
in planes of constant $z$.
A $z$--dependance of $\chi$ accounts for the finite thickness
of the quantum dielectric layer. 
Furthermore, matter and fields are assumed to be stationary, i.e. all
time derivatives vanish. 
We derive from Maxwell's equations (\ref{maxwell}) 
combined with the constitutive equations (\ref{constitution})
%%%%%%%%%%%%%%%%%%%%%%%%%%%%%%%%%%%%%%%%%%%%%%%%%%%%%%%%%%%%%
\begin{eqnarray}
\nabla {\bf H}&=&\varepsilon_0[
{\bf E} (\nabla\times\chi {\bf u})
- \chi {\bf u}(\nabla\times{\bf E})]
\nonumber\\
&=&\varepsilon_0{\bf E} (\nabla\times\chi {\bf u}) \,.
\label{h}
\end{eqnarray}
%%%%%%%%%%%%%%%%%%%%%%%%%%%%%%%%%%%%%%%%%%%%%%%%%%%%%%%%%%%%%
From the very beginning (\ref {la}) we have restricted our attention to
effects that occur in the lowest order of $v/c$. 
Hence we can replace the electric field ${\bf E}$ in Eq.\ (\ref{h}) 
by the ${\bf E}$--field in zeroth order (without R\"ontgen effect), 
i.e. by
%%%%%%%%%%%%%%%%%%%%%%%%%%%%%%%%%%%%%%%%%%%%%%%%%%%%%%%%%%%%
\begin{equation}\label{e0}
{\bf E}=E(z){\bf e}_z, \qquad E(z)=\frac{E_0}{1+\chi(z)}\,.
\end{equation}
%%%%%%%%%%%%%%%%%%%%%%%%%%%%%%%%%%%%%%%%%%%%%%%%%%%%%%%%%%%%%
The curl component of $\chi{\bf u}$ that stems from $\chi^{\prime}(z)$
is orthogonal on ${\bf e}_z$ and therefore on ${\bf E}$. 
Consequently,
%%%%%%%%%%%%%%%%%%%%%%%%%%%%%%%%%%%%%%%%%%%%%%%%%%%%%%%%%%%%
\begin{equation}
\nabla {\bf H}=\varepsilon_0\chi{\bf E}(\nabla\times{\bf u})\,.
\end{equation}
%%%%%%%%%%%%%%%%%%%%%%%%%%%%%%%%%%%%%%%%%%%%%%%%%%%%%%%%%%%%%
The source of the magnetic field is the curl of the velocity profile
projected onto the applied electric field. 
Two components contribute to the velocity, as is shown in Eq.~(\ref{u}).
One stems from the canonical momentum $\hbar\nabla S$ and the other
from the Poynting vector ${\bf E}\times{\bf B}$. 
Let us estimate the influence of the Poynting--vector contribution 
on the generation of ${\bf H}$. We find that
%%%%%%%%%%%%%%%%%%%%%%%%%%%%%%%%%%%%%%%%%%%%%%%%%%%%%%%%%%%%
\begin{eqnarray}
\left|\frac{\alpha}{m}{\bf E}\times({\bf E}\times{\bf B})\right|
&\le&
\frac{\alpha}{m}E^2|{\bf B}|
=\frac{\chi\varepsilon_0E^2}{mc^2|\psi|^2}|{\bf H}|
\nonumber \\
&\ll& |{\bf H}|\,,
\end{eqnarray}
%%%%%%%%%%%%%%%%%%%%%%%%%%%%%%%%%%%%%%%%%%%%%%%%%%%%%%%%%%%%%
because in realistic capacitors the electrostatic energy density 
$\varepsilon_0E^2/2$ is much smaller than $mc^2|\psi|^2$ 
(and otherwise pair production would occur).
Consequently, only the curl of $(\hbar /m) \nabla S$ 
generates the ${\bf H}$--field. 
However, the curl of a gradient uses to vanish, 
with the remarkable exception of vortices. 

A moving dielectric will not produce a magnetic field, 
except at localized vortices with the velocity profile 
(in cylindrical coordinates)
%%%%%%%%%%%%%%%%%%%%%%%%%%%%%%%%%%%%%%%%%%%%%%%%%%%%%%%%%%%%
\begin{equation}\label{vortex}
{\bf u}=\frac{\hbar}{m}\nabla S\,, \qquad 
S=-n\varphi\,,\qquad 
n \in \mathbb{Z}\,,
\end{equation}
%%%%%%%%%%%%%%%%%%%%%%%%%%%%%%%%%%%%%%%%%%%%%%%%%%%%%%%%%%%%%
because for a vortex (\ref{vortex}) the curl of ${\bf u}$ 
is proportional to a delta function. 
Vortices are quantized, since the wave functions 
$|\psi|\,\exp(iS)$ are supposed to be single--valued.
Let us calculate the ${\bf B}$ field of an isolated vortex. 
We derive from Maxwell's equations (\ref{maxwell}) and from 
the constitutive equations (\ref{constitution})
%%%%%%%%%%%%%%%%%%%%%%%%%%%%%%%%%%%%%%%%%%%%%%%%%%%%%%%%%%%%
\begin{equation}
\nabla\times {\bf B} = 
\frac{1}{c^2} \nabla\times({\bf E}\times\chi{\bf u})
\equiv \frac{1}{\varepsilon_0 c^2}{\bf j}_R\,.
\end{equation}
%%%%%%%%%%%%%%%%%%%%%%%%%%%%%%%%%%%%%%%%%%%%%%%%%%%%%%%%%%%%%
We utilize the continuity equation (\ref{c}) of the medium and the 
zeroth--order ${\bf E}$--field of Eq.\ (\ref{e0}) to arrive at
%%%%%%%%%%%%%%%%%%%%%%%%%%%%%%%%%%%%%%%%%%%%%%%%%%%%%%%%%%%%
\begin{equation}\label{b}
\nabla\times{\bf B}=
\frac{E_0}{c^2}\frac{\hbar n}{m \varrho}{\bf e}_{\varphi}
\frac{\partial}{\partial z}\left[\frac{\chi (z)}{1+\chi (z)}\right]\,.
\end{equation}
%%%%%%%%%%%%%%%%%%%%%%%%%%%%%%%%%%%%%%%%%%%%%%%%%%%%%%%%%%%%%
We see that the R\"ontgen current ${\bf j}_R$ is induced by the electric 
field ${\bf E}$ and is concentrated near the surfaces of the 
quantum dielectric where $\chi(z)$ varies significantly. 
We obtain the solution
%%%%%%%%%%%%%%%%%%%%%%%%%%%%%%%%%%%%%%%%%%%%%%%%%%%%%%%%%%%%
\begin{eqnarray}
{\bf B}&=&\int\limits_{-\infty}^{+\infty}f_n(z_0){\bf B}_0\,dz_0 \,,
\nonumber\\
{\bf B}_0&=&
-\nabla\frac{1}{\sqrt{\varrho^2+(z-z_0)^2}}-
\frac{2}{\varrho}\delta(z-z_0)
{\bf e}_{\varrho}\,,
\nonumber\\
f_n(z)&=&-\frac{\hbar n}{2 m c^2}\chi(z)E(z)\,,
\end{eqnarray}
%%%%%%%%%%%%%%%%%%%%%%%%%%%%%%%%%%%%%%%%%%%%%%%%%%%%%%%%%%%%%
because ${\bf B}_0$ satisfies
%%%%%%%%%%%%%%%%%%%%%%%%%%%%%%%%%%%%%%%%%%%%%%%%%%%%%%%%%%%%
\begin{eqnarray}
\nabla\times{\bf B}_0&=&
0-\frac{2}{\varrho}{\bf e}_{\varphi}\delta^{\prime} (z-z_0)\,,
\nonumber\\
\nabla{\bf B}_0&=&
4\pi \delta^3({\bf x}-z_0{\bf e}_z)
-2\cdot 2\pi \delta^3({\bf x}-z_0{\bf e}_z)=0\,.
\end{eqnarray}
%%%%%%%%%%%%%%%%%%%%%%%%%%%%%%%%%%%%%%%%%%%%%%%%%%%%%%%%%%%%%
Figure 2 shows the magnetic field lines that are bent by a vortex
in a charged capacitor. 
Inside the dielectric layer the vortex'
core attracts the field lines that leave the medium at a well
localized spot of roughly the thickness of the layer. 
In fact, we obtain in the limit of a thin layer
\begin{figure}[htbp]
  \begin{center}
    \epsfig{file=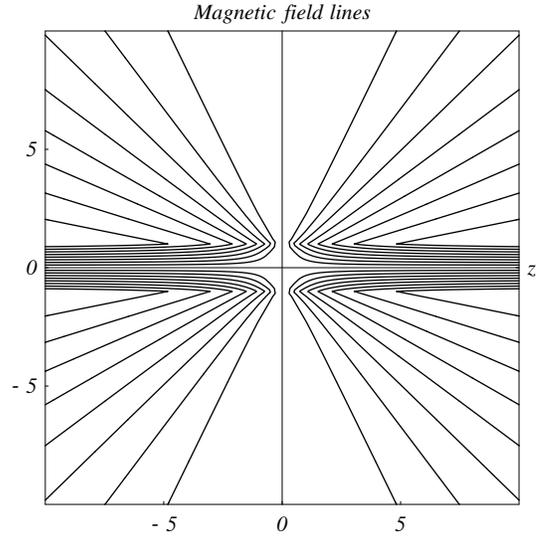,width=0.4\textwidth} 
    \caption{Quantum R\"ontgen effect.
    Consider a uniform layer of a quantum dielectric 
    (Bose--Einstein condensate) with a single vortex.
    The medium is polarized and generates a magnetic field.
    The field is rotationally symmetric, and the figure shows
    a two--dimensional cut of the magnetic field lines.
    Inside the layer (between $z=-1$ and $z=+1$
    in arbitrary units) field lines are attracted and 
    directed to the vortex' core (at the origin)
    where the lines leave the medium.
    In our simplified model the magnetic field lines close at infinity.}
    \label{fig:figure2}
  \end{center}
\end{figure}
\noindent
%%%%%%%%%%%%%%%%%%%%%%%%%%%%%%%%%%%%%%%%%%%%%%%%%%%%%%%%%%%%
\begin{equation}
{\bf B}\sim\frac{\Phi_n}{4\pi r^2}{\bf e}_r-
\frac{2 f_n(z)}{\varrho}{\bf e}_{\varrho}\,,
\end{equation}
%%%%%%%%%%%%%%%%%%%%%%%%%%%%%%%%%%%%%%%%%%%%%%%%%%%%%%%%%%%%%
written using spherical coordinates with $r^2=\varrho^2+z^2$. 
Outside the medium the vortex appears as a three--dimensional
magnetic monopole with a magnetic flux of
 %%%%%%%%%%%%%%%%%%%%%%%%%%%%%%%%%%%%%%%%%%%%%%%%%%%%%%%%%%%%
\begin{equation}
\Phi_n=4\pi \int\limits^{+\infty}_{-\infty}f_n(z)dz=
-\frac{hn}{mc^2}\int \chi {\bf E}\,d{\bf r}\,.
\end{equation}
%%%%%%%%%%%%%%%%%%%%%%%%%%%%%%%%%%%%%%%%%%%%%%%%%%%%%%%%%%%%
For a uniform layer with a constant susceptibility $\chi$ inside and 
a vanishing $\chi$ outside we arrive at the simple result (\ref{result}) 
that was mentioned in the introductory part of this paper.
Inside the medium the ${\bf B}$ field appears as a two--dimensional
monopole field that supplies the 3D monopole with field lines.

{\it Can one measure the effect?}---
Despite the breathtaking progress in the Bose-Einstein condensation of
alkali atoms since the pioneering breakthrough \cite{Pioneers}, 
superfluid helium is probably still the best candidate for a 
quantum dielectric medium.
Helium is light with a rest mass of roughly four proton masses and the
liquid is relatively dense with a susceptibility $\chi=0.052$. 
Using these numbers we obtain a quantized flux $\Phi_n$
per voltage $U$ of $n\cdot 5.7\cdot 10^{-26}\,[{\rm Tm^2/V}]$. 
The flux is quite small yet the magnetic field is most probably 
measurable using nano--fabricated SQUIDs, 
taking advantage of the fact that the field stems from an 
extremely well localized spot.
In fact, for a nanometer--size helium film we obtain a
magnetic field (25) of a few femtotesla per applied volt at
a single vortex with $n=1$, and this field seems to be measurable.
One could also think of using much thinner helium layers
that will lead to stronger magnetic fields.
A helium film can be 0.1 monolayers thick and still be
superfluid (where a monolayer has a thickness of
$3.6\cdot 10^{-10}\, {\rm m}$).
The quantized R\"ontgen effect could be applied to detect
vortices in flowing helium films and to study their interactions.
The generated magnetic field may serve as a sensor for vorticity.
As we have seen, vortices of quantum dielectrics give rise
to interesting magnetic fields --- 3D monopole fields 
outside and 2D monopole fields inside the layer, 
and we find it worthwhile to study these fields experimentally.
Superfluid helium is probably the best material to observe the
quantized R\"ontgen effect with existing techniques,
but it is also conceivable to employ optically trapped
alkali or other Bose--Einstein condensates.

{\it Summary.}---
A moving quantum dielectric sandwiched between two charged capacitor
plates will not produce a magnetic field, except at vortices. 
Here the generated field outside the medium
appears as the field of a magnetic monopole 
with the quantized flux (\ref{result}) 
and a magnetic charge that is localized in the vortex' core.

Our paper complements the recent effort 
\cite{Wilkens,China,Discussion,Experiment} 
in employing the R\"ontgen interaction of traveling dipoles
for novel topological effects. In this case the electromagnetic field
forms a vortex of the Poynting vector and acts on moving atoms. 
In our case, the atomic condensate acts on the fields in the same way 
as a moving dielectric medium. Quantum mechanics, however, 
restricts the properties of the medium. 
We believe that our approach of regarding a Bose--Einstein 
condensate as a dielectric medium will open a new research road in the
fascinating field of quantum gases. 
Subtle quantum effects may lead to new surprises.

We are grateful to
E. Anderson,
P.~J. Bardroff,
M.~V. Berry,
M. Fontanelle,
D.~B. Haviland,
M. Krusius,
P. \"Ohberg,
M. Revzen,
W. Schleich,
S. Stenholm,
M. Wilkens,
and
A.~F.~G. Wyatt
for helpful discussions.
U.~L. thanks the Alexander von Humboldt Foundation
and the G\"oran Gustafsson Stiftelse for support.
P.~P. was supported by the 
research consortium {\it Quantum Gases} of the
Deutsche Forschungsgemeinschaft.

\end{document}